\documentclass[10pt,aps,prd,twocolumn,showpacs,amsmath,nofootinbib]{revtex4-2}

\usepackage[utf8]{inputenc}
\usepackage[american]{babel}
\usepackage[a4paper,margin=1.6cm]{geometry}

\usepackage{amsmath,amssymb,mathtools}
\usepackage{booktabs}
\usepackage{graphicx}
\usepackage[labelfont=bf,font+=smaller,justification=centerlast]{caption}
\usepackage[labelfont=normal,font=normal,font=small]{subcaption}
\usepackage{slashed}

\DeclareTextSymbolDefault{\softl}{T1}
\DeclareTextCommand{\softl}{T1}{\v{l}}
\DeclareUnicodeCharacter{013E}{\softl}

\usepackage{hyperref}
\usepackage{cleveref}
\renewcommand{\cref}[1]{\Cref{#1}}

\Crefname{equation}{Eq.}{Eqs.}
\Crefname{figure}{Fig.}{Figs.}
\Crefname{tabular}{Tab.}{Tabs.}
\Crefname{section}{Sec.}{Secs.}

\hypersetup{%
colorlinks,%
citecolor=blue,urlcolor=blue,linkcolor=blue,%
hypertexnames=false%
}

\newcommand\pigg{\pi^0\to\gamma\gamma}
\newcommand\piee{\pi^0\to e^+e^-}
\newcommand\pieeg{\pi^0\to e^+e^-\gamma}
\newcommand\pieegx{\pi^0\to e^+e^-(\gamma)}

\def\Mpi{M_{\pi^0}}

\def\chit{\widetilde\chi}
\def\chir{\chi^{(\mathrm{r})}}
\def\xcut{x_\text{cut}}

\def\deltapm{\delta_\pm}
\def\deltap{\delta_+}
\def\deltam{\delta_-}
\def\deltatotal{\delta}
\def\deltavirt{\delta}
\def\deltaxcut{\delta}
\def\deltax{\hat\delta}
\def\Deltaxcut{\Delta}
\def\Deltax{\hat\Delta}
\def\Deltasoft{\Delta}
\def\DeltaZ{\Delta_Z}

\allowdisplaybreaks[0]

\hypersetup{%
pdftitle={The neutral-pion decay into electron-positron pair: A review and update},%
pdfauthor={T. Husek}%
}

\begin{document}

\title{Neutral-pion decay into an electron--positron pair: A review and update}
\author{Tom\'{a}\v{s} Husek}
\email{tomas.husek@matfyz.cuni.cz; t.husek@bham.ac.uk}
\affiliation{Institute of Particle and Nuclear Physics, Charles University, Prague, Czech Republic}
\affiliation{School of Physics and Astronomy, University of Birmingham, Edgbaston, Birmingham, B15 2TT, UK}
\date{August 15, 2024}

\begin{abstract}
This work aims to review several aspects of the current status of the rare $\pi^0\to e^+e^-$ decay. A particular emphasis is made on radiative corrections and detailed interpretation of related quantities, some numbers appearing in the literature are updated, and the connection with the Dalitz decay, $\pi^0\to e^+e^-\gamma$, is discussed. This comes timely as it is aligned with an announcement of a preliminary result of a new branching-ratio measurement done by the NA62 Collaboration, which brings new light into an earlier-reported discrepancy between the Standard Model prediction and the (until-recently latest) precise KTeV result.
\end{abstract}


\maketitle

\section{Introduction}

With respect to the branching ratio (BR) of the radiative decay $\pigg$, the rare decay $\piee$ is loop- and helicity-suppressed by eight orders of magnitude.
One might thus speculate that this channel is sensitive to possible measurable effects of new physics.
Indeed, this seemed to be the case when the KTeV Collaboration published the results of their analysis on the BR precise measurement in 2007~\cite{KTeV:2006pwx}.
The direct subsequent comparison to the Standard Model (SM) prediction was then interpreted as a 3.3\,$\sigma$ discrepancy~\cite{Dorokhov:2007bd}.
Many works followed trying to explain this difference both within (via models and predictions for the pion transition form factor) and outside the SM (introducing various models including exotic particles).

The NA62 Collaboration presented the preliminary result of 2017--2018-dataset analysis at the ``Rencontres de Moriond 2024'' conference, and the publication of the final result is under preparation.
The overall uncertainty is at the same level as the KTeV measurement, but the central value shifted significantly: It is smaller, perfectly consistent with theoretical expectations based on the SM considerations.
Finally, a more recent and larger NA62 dataset is available, and eventually, thousands of $\piee$ signal events are planned to be analyzed.

It has been argued in the past that improper handling of radiative corrections (RCs) might have been an essential source of the discrepancy.
It is believed that the most significant contribution in this regard was made by an explicit calculation of the (two-loop) virtual radiative corrections~\cite{Vasko:2011pi} that brought the persisting tension down to $2\,\sigma$ level~\cite{Husek:2014tna,Husek:2015wta}: Until then, only results based on leading-log approximation were available~\cite{Bergstrom:1982wk,Dorokhov:2008qn} that --- in the light of the subsequent exact calculation \cite{Vasko:2011pi} --- did not reproduce well cancellations among contributing terms and thus overestimated the size of the correction.

The (latest) NA62 analysis makes use of the available complete set of the related next-to-leading-order (NLO) QED radiative corrections~\cite{Vasko:2011pi,Husek:2014tna,Husek:2015sma,Husek:2018qdx} already at Monte Carlo (MC) level, including radiative modes with an extra photon.
This leads to better control over acceptance and gains better data--MC agreement.
It is, therefore, instructive to revisit the corrections worked out to date and revise, discuss, and summarize them in one spot.
This is the aim of the present work, which in detail describes the relation between the theoretical (\cref{sec:LO}) and experimental (\cref{sec:experiment}) observables, updates the critical NLO QED correction $\deltap(\xcut)$ that relates these, calculates the overall correction $\deltatotal$ (\cref{sec:NLO}), tabulates a newly defined correction $\deltam(\xcut)$ for sample cut-off values (\cref{tab:xcut}), derives other related useful experimental quantities or ratios and discusses the latest measurements (\cref{sec:measurements}), dives into some interesting properties of the $\pieegx$ amplitude in various limits (\cref{sec:BSm0} and \cref{sec:BS}), and discusses its relation with the Dalitz decay radiative corrections (\cref{sec:1gIR}).
The findings are accompanied by some instructive plots.

\section{Leading order in the ChPT expansion}
\label{sec:LO}

\begin{figure}[htb]
\includegraphics[width=\columnwidth]{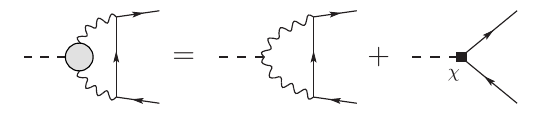}
\caption{
    Leading-order contribution to $\piee$ in the QED expansion and its representation in terms of the leading order in the Chiral Perturbation Theory.
    The gray blob represents the $\pi^0\to\gamma^*\gamma^*$ transition form factor.
}
\label{fig:LO}
\end{figure}

The leading-order (LO) matrix element in the QED expansion for $\piee$ is represented by the one-loop diagram shown on the left-hand side of the graphical equation in \cref{fig:LO}, and it can be written in terms of a nonperturbative quantity, the doubly virtual neutral-pion transition form factor (TFF).
The latter can be further expanded within Chiral Perturbation Theory (ChPT) as a pointlike interaction accompanied by a compensatory counterterm (see the right-hand side of \cref{fig:LO}) mimicking the remaining high-energy contribution.
The LO in this double expansion contains only a single hadronic parameter, the finite part of a constant named $\chi$ (a combination of low-energy constants appearing in the ChPT counterterm Lagrangian with dynamical leptons~\cite{Savage:1992ac}); the higher-order corrections in the ChPT expansion are expected to be tiny, as demonstrated in the leading-log approximation in Ref.~\cite{Husek:2014tna}.
Accounting also for a small contribution from the weak interaction, related BR prediction that does not take into account any final-state radiation can thus be written in a model-independent way,%
\footnote{
The first line of \cref{eq:pieeLO} alone represents the contribution of the absorptive part of the amplitude from on-shell photons, providing thus the unitarity lower bound~\cite{Drell:1959,Berman:1960zz}, leading to $B\bigl(\piee\bigr)\gtrsim4.69\times10^{-8}$.
}
\begin{multline}
\frac{B\bigl(\piee\bigr)}{B\bigl(\pi^0\to\gamma\gamma\bigr)}
=\frac\beta2\biggl(\frac\alpha\pi\frac{m_e}{\Mpi}\biggr)^2
\biggl\{
\bigg[\frac\pi{\beta}\log z\bigg]^2\\
+\biggl[\chit(\mu)+3\log\frac{m_e^2}{\mu^2}+\DeltaZ+\frac1{\beta}\biggl(\operatorname{Li}_2(-z)-\operatorname{Li}_2\frac1{(-z)}\biggr)\biggr]^2
\biggr\}\,,
\label{eq:pieeLO}
\end{multline}
with $\beta\equiv\sqrt{1-\nu^2}\approx1$, $\nu\equiv{2m_e}/{\Mpi}\approx0.757\%$, and $z=\frac{1-\beta}{1+\beta}\simeq\frac{\nu^2}4$, with $Z$-boson-exchange contribution $\DeltaZ=-4\frac{\pi^2}{\alpha^2}{\sqrt{2}F_\pi^2 G_\text{F}}\approx-0.10$~\cite{Masjuan:2015cjl,Hoferichter:2021lct}, and where I have introduced an artificial (although naturally appearing) combination
\begin{equation}
\chit(\mu)
\equiv2\bigl[\chir(\mu)-\tfrac52\bigr]\,.
\label{eq:chit}
\end{equation}
Since relevant theoretical models typically suggest (being rather conservative) $\chir(770\,\mathrm{MeV})\approx2\text{--}3$ (see also \cref{tab:chi}), one conveniently has $\chit\equiv\chit(770\,\mathrm{MeV})\in(-1,1)$, in terms of which the LO BR reads
\begin{equation}
B\bigl(\piee\bigr)
\approx\bigl(6.193+0.152\chit+0.004\chit^2\bigr)\times10^{-8}\,,
\label{eq:Bpi0ee}
\end{equation}
or simply $B\bigl(\piee\bigr)\approx\bigl(6.19+0.15\chit\bigr)\times10^{-8}$.
\setlength{\tabcolsep}{10pt}
\begin{table*}[th]
\begin{ruledtabular}
\begin{tabular}{l | l l | l}
     & $B\bigl(\pieegx,\,x>0.95\bigr)\;\bigl[10^{-8}\big]$ & $B\bigl(\piee\bigr)\;\bigl[10^{-8}\big]$ & $\chir(770\,\text{MeV})$\\
    \midrule
    PDG (2023) \cite{Workman:2022ynf} & 6.46(33) & \\
    KTeV (2007) \cite{KTeV:2006pwx} & 6.44(33) & 7.48(38) & 6.0(1.0)\footnotemark[2] \\
    KTeV + RCs of Refs.~\cite{Vasko:2011pi,Husek:2014tna} & ~~~$\hookrightarrow$ & 6.84(35)\footnotemark[2] & 4.5(1.0) \\
    \midrule
    NA62 preliminary (2024) & 5.86(37) & 6.22(39) & 2.5(1.3)\footnotemark[2]\\
    \midrule
    Knecht et al. (1999) \cite{Knecht:1999gb} & 5.8(3)\footnotemark[1] & 6.2(3) & 2.2(9) \\
    Dorokhov and Ivanov (2007) \cite{Dorokhov:2007bd} & 5.85(10)\footnotemark[1] & 6.23(9) & 2.6(3) \\
    Husek and Leupold (2015) \cite{Husek:2015wta} & 5.75(7)\footnotemark[1] & 6.12(6)\footnotemark[2] & 2.2(2) \\
    Hoferichter et al. (2022) \cite{Hoferichter:2021lct} & 5.87(4)\footnotemark[1] & 6.25(3) & 2.69(10) \\
\end{tabular}
\caption[A comparison of latest experimental and some theoretical results.
    The quoted values lacking decoration come directly from the mentioned references.]
    {A comparison of latest experimental and some theoretical results.
    The quoted values lacking decoration come directly from the mentioned references.
    \footnotetext[1]{
        For these values, the correction $\deltap(0.95)=-6.1(2)\%$ [\cref{eq:deltap_95_new}] was used.}
    \footnotetext[2]{
        These values are obtained based on relation \eqref{eq:Bpi0ee}.}
}
\label{tab:chi}
\end{ruledtabular}
\end{table*}
Written in this way, this quantity is more transparent than an equivalent expression using $\chir(770\,\mathrm{MeV})$, as it allows one to immediately see the approximate central value of the SM prediction, its (conservative) $1\,\sigma$ band covering considered models, and it also provides the possibility for direct extraction of $\chir$ when matched to particular predictions or measurements.
Moreover, this redefinition also works for derived quantities, like the subamplitudes of the $\pieegx$ amplitude, for which the combination in \cref{eq:chit} also appears; cf.~Appendix A of Ref.~\cite{Husek:2014tna}.

This exclusive, purely theoretical BR cannot be directly accessed experimentally, and --- for the theory--experiment comparison --- QED corrections have to be used to subtract the radiative effects.

\section{Experimental quantities}
\label{sec:experiment}

\begin{figure*}[tb]
\centering
\begin{subfigure}{0.47\textwidth}
    \centering
    \includegraphics[width=\linewidth]{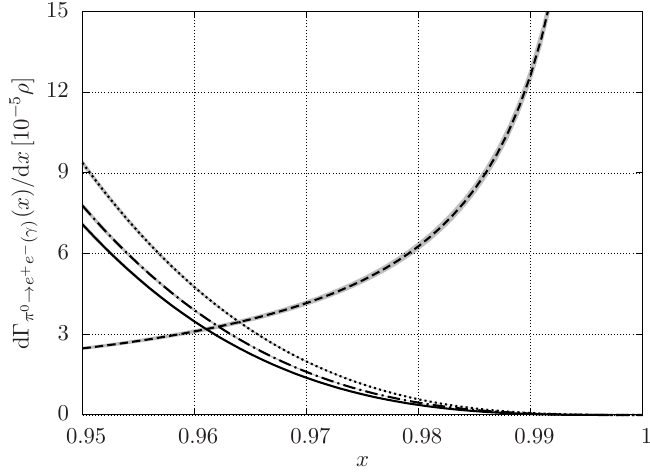}
    \caption{}
\label{fig:Gamma_pi0eeg_95}
\end{subfigure}
\hfill
\begin{subfigure}{0.47\textwidth}
    \centering
    \includegraphics[width=\linewidth]{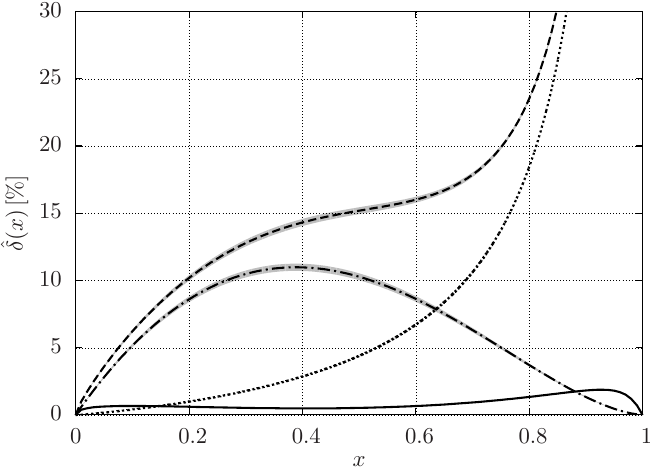}
    \caption{}
\label{fig:Gamma_pi0eeg}
\end{subfigure}
\caption{
    The one-fold differential decay width of $\pieegx$ with respect to $x$.
    In the panel (a), I use the units of $\rho\equiv\displaystyle\frac\alpha\pi\Gamma_{\pi^0\to\gamma\gamma}^\text{LO}$, with $\Gamma_{\pi^0\to\gamma\gamma}^\text{LO}=\displaystyle\frac{\Mpi^3}{64\pi}\biggl(\frac{e^2}{4\pi^2F_\pi}\biggr)^2=7.76(2)\,\text{eV}$;
    in the panel (b), it is expressed in the units of $\Gamma_{\piee}$, i.e., as the radiative correction $\deltax(x)$, in percent.
    In the panel (a), the $\pieegx$ decay width (dashed line) is compared with the Dalitz decay at LO (dotted line) and at NLO without (dash-dotted line) and with (solid line) the 1$\gamma$IR correction in the region $x>0.95$.
    In the panel (b), there is a comparison of the exact NLO result (dashed line) with the limit $m_e\to0$ (dash--dotted line) and the soft-photon limit in the sense of \cref{eq:deltax} (dotted line).
    The sum of these two limiting cases represents the exact result very well, and the difference of the exact result and this sum is plotted as a solid line.
    Note that the dashed line represents the same quantity in both panels, only different units are used.
    For $x>0.95$ [panel (a)], the soft-photon limit \eqref{eq:deltax} basically overlaps with the exact result and is not plotted.
    The shaded area corresponds to the uncertainty on $\chir$ and, in the case of the Dalitz-decay widths, TFF slope $a_\pi^\text{univ}=3.55(70)\%$~\cite{Husek:2018qdx}.
}
\label{fig:gamma}
\end{figure*}

Due to the inevitable natural presence of additional photons in processes featuring charged particles, a well-defined (observable) experimental quantity in our case would be
\begin{multline}
B\bigl(\pieegx,\,x>\xcut\bigr)\\
=\big[1+\deltaxcut(\xcut)\big]B\bigl(\piee\bigr)\,.
\label{eq:Bpi0eeg}
\end{multline}
Such a BR measured in a real-world experiment is thus inclusive of (final-state) soft photons, which is a fortunate consequence of limited detector sensitivity to such electromagnetic quanta.
The quantity \eqref{eq:Bpi0eeg} is defined in terms of a cut-off $\xcut$ on the kinematical variable
\begin{equation}
x
\equiv\bigl(p_{e^+}+p_{e^-}\bigr)^2/\Mpi^2\,,
\label{eq:x}
\end{equation}
the normalized electron--positron invariant mass squared.
Ideally, $\xcut$ should be chosen large enough to suppress the effects of a competing process with very different dynamics but the same final state, the neutral-pion Dalitz decay, $\pieeg$, which in turn peaks at low $x$; see also \cref{fig:Gamma_pi0eeg_95}.

There are several options for how to define further what actually the measurement of $B\bigl(\pieegx,\,x>\xcut\bigr)$ might mean.
Without closer specification, this quantity naturally has --- beyond the theoretical ``no-radiation'' $\piee$ part --- several components: the $\piee$ bremsstrahlung (BS), the Dalitz decay, and their interference.
It is thus essential to know what exactly is (or is not) included in a particular result for this measured quantity.

Specifically, in the case of the latest NA62 measurement, the Dalitz decay and its interference with the rare-decay BS have been subtracted, as this interference [also referred to as the one-photon-irreducible ($1\gamma$IR) correction] is part of the radiative corrections to the Dalitz decay in the NA62 MC;
note that this was not the case for the KTeV measurement, for which only the classical RCs due to Mikaelian and Smith~\cite{Mikaelian:1972yg} were considered (the reasons are mentioned in \cref{sec:1gIR}) and subtracted.
This means that, regarding the NA62 result, $B\bigl(\pieegx,\,x>\xcut\bigr)$ has a clear meaning of the BR related solely to $\piee$ and its BS.
It can thus be related to $B\bigl(\piee\bigr)$ using \cref{eq:Bpi0eeg}, taking as input the corrections addressed in Refs.~\cite{Vasko:2011pi,Husek:2014tna} and in this work.

In particular, knowing the factor $\big[1+\deltaxcut(\xcut)\big]$, $B\bigl(\piee\bigr)$ and $B\bigl(\pieegx,\,x>\xcut\bigr)$ are equivalent and can thus both be used to extract $\chir$, as long as their meanings are precisely defined (in terms of what is subtracted and what is included in these quantities, as discussed above).

\section{Radiative corrections}
\label{sec:NLO}

\begin{figure}[tb]
\centering
    \includegraphics[width=\linewidth]{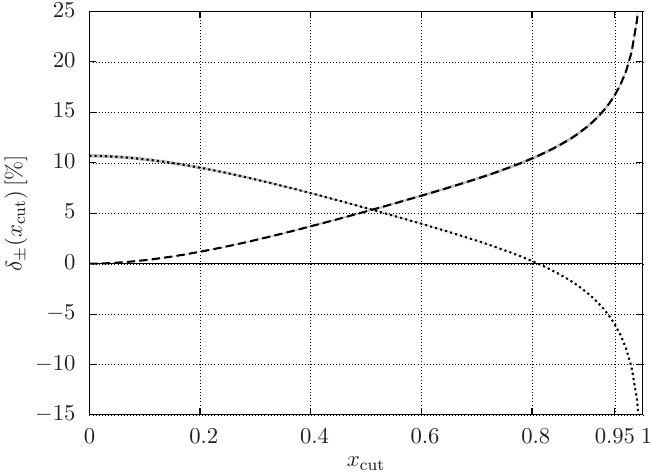}
    \caption{
    Radiative corrections to $\piee$.
    The correction $\deltam(\xcut)$ of \cref{eq:deltapm} (dashed line) is accompanied with $\deltap(\xcut)$ (dotted line).
    The sum of these is a constant $\deltatotal$.
    }
\label{fig:delta_pm}
\end{figure}

It is convenient to introduce the following quantities:
\begin{equation}
\deltapm(\xcut)
\equiv\frac{\Gamma^\mathrm{NLO}\bigl(\pieegx,\,x\gtrless \xcut\big)}{\Gamma^\mathrm{LO}\bigl(\piee\big)}\,.
\label{eq:deltapm}
\end{equation}
For any $\nu^2\le\xcut<1$, the total NLO correction is then a constant
\begin{equation}
\deltatotal
\equiv\deltap(\nu^2)
=\deltam(\xcut)+\deltap(\xcut)\,;
\label{eq:delta_gen}
\end{equation}
see \cref{fig:delta_pm} for $\deltam(\xcut)$ and \cref{tab:xcut} for its precise sample values.
\begin{table*}[bt]
\begin{ruledtabular}
\begin{tabular}{c r c | c r c | c r c | c r}
$\xcut$ & $\deltam(\xcut)\,[\%]$\hspace{1mm} &&
$\xcut$ & $\deltam(\xcut)\,[\%]$\hspace{-1mm} &&
$\xcut$ & $\deltam(\xcut)\,[\%]$\hspace{-1mm} &&
$\xcut$ & $\deltam(\xcut)\,[\%]$ \\
\midrule
0.05 & 0.09973(130) &&
0.35 & 3.000(57)\hphantom{0} &&
0.65 & 7.552(130) &&
0.95\hphantom{0} & 16.74(17)~~ \\
0.10 & 0.3468(57)\hphantom{00} &&
0.40 & 3.700(71)\hphantom{0} &&
0.70 & 8.413(140) &&
0.96\hphantom{0} & 17.76(17)~~ \\
0.15 & 0.7137(130)\hphantom{0} &&
0.45 & 4.429(84)\hphantom{0} &&
0.75 & 9.352(150) &&
0.97\hphantom{0} & 19.08(17)~~ \\
0.20 & 1.180(22)\hphantom{000} &&
0.50 & 5.179(97)\hphantom{0} &&
0.80 & 10.43(15)\hphantom{00} &&
0.98\hphantom{0} & 20.96(17)~~ \\
0.25 & 1.726(33)\hphantom{000} &&
0.55 & 5.947(110) &&
0.85 & 11.76(16)\hphantom{00} &&
0.99\hphantom{0} & 24.19(17)~~ \\
0.30 & 2.338(45)\hphantom{000} &&
0.60 & 6.736(120) &&
0.90 & 13.59(16)\hphantom{00} &&
0.995 & 27.43(17)~~ \\
\end{tabular}
\end{ruledtabular}
\caption{
The values of $\deltam(\xcut)$ from \cref{eq:deltapm} for chosen sample values of $\xcut$.
To be suitable for interpolation, more digits are displayed.
The quoted uncertainties are dominated by the TFF knowledge (for its value, I assume $\chi_\text{univ}^{(\mathrm{r})}(770\,\text{MeV})=2.5(5)$, corresponding to $\chit=0(1)$).
As the dependence of $\deltam(\xcut)$ on $\chi_\text{univ}^{(\mathrm{r})}$ is linear to a very good approximation, one can estimate well the correction for a particular $\chir$ value, taking into account that for larger $\chir$ the value of $\deltam(\xcut)$ gets smaller.
For instance, assuming $\chir=2.7$, $\deltam(0.95)=16.74-\frac250.17=16.67$.
In other words, $\deltam(0.95)=16.74(17)_\chi\equiv16.74-0.17\chit$.
}
\label{tab:xcut}
\end{table*}%

The correction in \cref{eq:Bpi0eeg} can be approximated by the NLO computation as
\begin{equation}
\deltaxcut(\xcut)
\simeq\deltap(\xcut)\,.
\end{equation}
For the standard choice $\xcut=0.95$ (due to KTeV and its predecessor), the value for the NLO radiative correction accounting for the events with one extra soft photon (in the vicinity of the region $x\simeq1$) has been changing.
First estimates~\cite{Bergstrom:1982wk,Dorokhov:2008qn} led to rather large values, $\deltap(0.95)\approx-13.5\%$; see also Ref.~\cite{Vasko:2011pi} for more details.
However, the most recent exact NLO calculations found
\begin{equation}
\begin{aligned}
\deltap(0.95)
&=\deltap^\text{(soft+virt.)}(0.95)+\Deltaxcut^\text{BS}(0.95)\\
&=-5.8(2)\%~\text{\cite{Vasko:2011pi}}+0.30(1)_\chi\%~\text{\cite{Husek:2014tna}}\\
&=-5.5(2)\%\,.
\end{aligned}
\label{eq:deltap_95}
\end{equation}
I would like to argue that, unfortunately, this number is not the final answer either.

The quantity $\deltap^\text{(soft+virt.)}(\xcut)$ related to the first figure on the middle line of \cref{eq:deltap_95}, $-5.8(2)\%$, is unaffected by the proposed update, and I only bring out its $\chit$ dependence.
It originates in the NLO calculation that considers the BS contribution in soft-photon approximation, with the result presented in Eq.~(8.4) of Ref.~\cite{Vasko:2011pi}, and has the following decomposition:
\begin{equation}
\begin{aligned}
\deltap^\text{\cite{Vasko:2011pi}}(\xcut)
&\equiv\deltap^\text{(soft+virt.)}(\xcut)\\
&=\deltap^\text{(soft)}(\xcut)+\deltavirt^\text{virt}\\
&=c_1\log(1-\xcut)+c_2+\deltavirt^\text{virt}\,.
\end{aligned}
\label{eq:delta$x_cut$soft}
\end{equation}

The expressions for $c_1$ and $c_2$ are known precisely (they do not depend on the low-energy constants under consideration) and can be written as compactly as
\begin{equation}
\begin{multlined}
c_1
=\frac\alpha\pi
(-2)\bigl(1+\widetilde\beta\log z\bigr)\,,\\
c_2
=\frac\alpha\pi
\biggl\{
\bigl(
1+\widetilde\beta\log z
\bigr)
\biggl(
\log\frac{m_e^2}{\Mpi^2}
-\bar\gamma
\biggr)
\hfill\\
-\frac1\beta\log z
-\widetilde\beta
\biggl[
\frac12\log^2z+2\operatorname{Li}_2(1-z)
\biggr]
\biggr\}\,,
\end{multlined}
\label{eq:ci}
\end{equation}
with $\widetilde\beta\equiv\frac{1+\beta^2}{2\beta}
\simeq1+\mathcal{O}(\nu^4)$
and $\bar\gamma\equiv\gamma_\mathrm{E}-\log4\pi$.
Numerically, $c_1\approx4.717\%$ and $c_2\approx9.075\%$.

The virtual correction in the given scheme is numerically estimated to be $\deltavirt^\text{virt}=-0.8(2)\%$, as presented in Eq.~(8.3) of Ref.~\cite{Vasko:2011pi}.
Its uncertainty has two major components: knowledge of $\chir$ \{in Ref.~\cite{Vasko:2011pi}, $\chir(770\,\text{MeV})=2.2(9)$ was used\} and an estimate on an unknown NLO counterterm coupling involved, $\xi^{(\text{r})}(770\,\text{MeV})=0\pm5.5$; see also end of Sec.~6.4 of Ref.~\cite{Vasko:2011pi}.
Considering the precise values for $c_1$ and $c_2$ and examining Table 3 and Figure 10 of Ref.~\cite{Vasko:2011pi}, I find
\begin{equation}
\deltavirt^\text{virt}
=\bigl[-0.82(7)_\xi-0.08\chit\bigr]\%\,.
\end{equation}

As the expression \eqref{eq:delta$x_cut$soft} represents the exact result best in the soft-photon region, i.e., for $\xcut\to1$, to reach a more precise answer further away from that point, one might either include a compensatory term $\Deltaxcut^\text{BS}(\xcut)$ \{the strategy adopted in Ref.~\cite{Husek:2014tna} and indicated in \cref{eq:deltap_95}\}, or calculate $\deltap(\xcut)$ in a spirit of repeated use of \cref{eq:delta_gen} as
\begin{equation}
\begin{aligned}
\deltap(\xcut)
&=\deltatotal-\deltam(\xcut)\\
&=\bigl[\deltam(1-\epsilon)+\deltap(1-\epsilon)\bigr]-\deltam(\xcut)\,,
\end{aligned}
\label{eq:deltacut$new$}
\end{equation}
with $0<\epsilon\ll1$, e.g., $\epsilon\lesssim10^{-3}$.
That is, using $\deltatotal$ obtained as a sum of $\deltam(1-\epsilon)$, an exact-to-NLO and IR-finite piece, and $\deltap(1-\epsilon)\simeq\deltap^\text{(soft)}(1-\epsilon)$ that can be rather precisely evaluated in the soft-photon limit, i.e., using \cref{eq:delta$x_cut$soft}.
Hence,
\begin{equation}
\begin{aligned}
\deltatotal
&=\bigl[11.494-0.168\chit\bigr]\%+\deltavirt^\text{virt}\\
&=\bigl[10.67(7)_\xi-0.25\chit\bigr]\%\\
&=10.7(1)_\xi(2)_\chi\%\,,
\end{aligned}
\label{eq:delta}
\end{equation}
and as $\deltam(0.95)=\bigl[16.737-0.166\chit\bigr]\%$ (cf.~\cref{tab:xcut} and its caption), one finally obtains
\begin{equation}
\begin{aligned}
\deltap(0.95)
&=-5.243(2)_\chi\%+\deltavirt^\text{virt}\\
&=\bigl[-6.06(7)_\xi-0.08\chit\bigr]\%\\
&=-6.1(2)\%\,.
\end{aligned}
\label{eq:deltap_95_new}
\end{equation}

\begin{figure*}[tb]
\centering
\begin{subfigure}{0.48\textwidth}
    \centering
    \includegraphics[width=\linewidth]{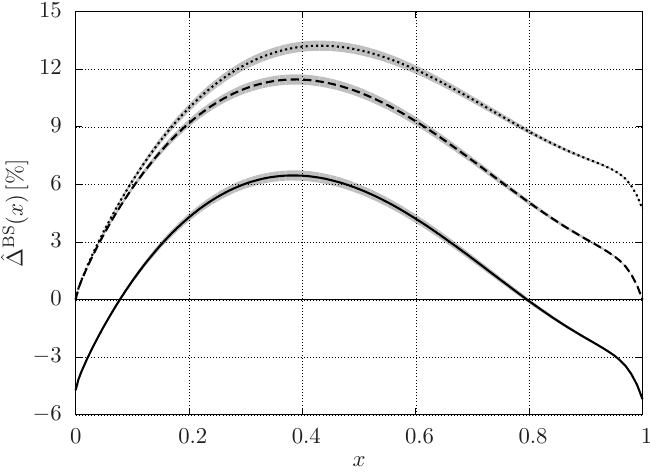}
\caption{}
\label{fig:DeltaBSx}
\end{subfigure}
\hfill
\begin{subfigure}{0.48\textwidth}
    \centering
    \includegraphics[width=\linewidth]{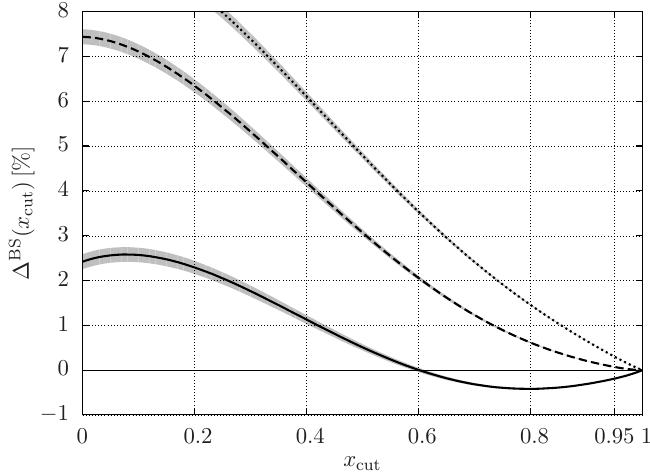}
\caption{}
\label{fig:DeltaBSxcut}
\end{subfigure}
\caption{
    The compensatory term $\Deltaxcut^\text{BS}$; see also \cref{sec:BS}.
    In the panel (a), there is the (IR-finite) difference [integrand of \cref{eq:DeltaBsxcut}] $\Deltax^\text{BS}(x)=\deltax^\text{BS}(x)-\deltax^\text{(soft)}(x)$ of the exact NLO correction $\deltax^\text{BS}(x)$ and several soft-photon approaches: employing $\deltax^\text{(soft)}(x)$ of \cref{eq:deltax} (dashed line), $\deltax_{x\to1}^\text{(soft)}(x)$ of \cref{eq:deltax1} (solid line), and $x\deltax^\text{(soft)}(x)$ as used in Ref.~\cite{Husek:2014tna} (dotted line).
    This plot replaces the one in Figure~5 of Ref.~\cite{Husek:2014tna}.
    In the panel (b), one can see the integrals over $x\in(\xcut,1)$ of the curves in the panel (a).
    This plot replaces the one in Figure~6 of Ref.~\cite{Husek:2014tna}:
    The original curve is represented by a dotted line, and the new result that allows for negative values is plotted as a solid line.
    The dashed line corresponds to the most natural approximation $\deltax^\text{(soft)}(x)$.
    The shaded area corresponds to the variation of $\chir$ in the range $\chir(770\,\text{MeV})\in(2,3)$.
}
\label{fig:DeltaBS}
\end{figure*}

Now, why the results in \cref{eq:deltap_95,eq:deltap_95_new} are so different when they should be the same?
It turns out that, unfortunately, in Ref.~\cite{Husek:2014tna}, a different soft-photon-limit convention was used with respect to Ref.~\cite{Vasko:2011pi}, and the therein-evaluated $\Deltaxcut_\text{\cite{Husek:2014tna}}^\text{BS}(0.95)=0.30(1)_\chi\%$ thus does not have the desired meaning.
Instead, to be on par with the soft-photon result of Ref.~\cite{Vasko:2011pi}, the compensatory term should read $\Deltaxcut^\text{BS}(0.95)=-0.18(1)_\chi\%$; see also \cref{sec:BS} for more technical details and \cref{fig:DeltaBS} for comparison.
The original message of Ref.~\cite{Husek:2014tna}, however, holds:
For $\xcut=0.95$, the soft-photon approximation featuring in results of Ref.~\cite{Vasko:2011pi} is (given the uncertainty) more or less sufficient.

The use of the updated value for $\Deltaxcut^\text{BS}(0.95)$ in \cref{eq:deltap_95} leads to a result consistent with the approach chosen and preferred here [\cref{eq:deltacut$new$} and below], i.e.,
\begin{equation}
\deltam(\xcut)
+\deltap^\text{\cite{Vasko:2011pi}}(\xcut)
+\Deltaxcut^\text{BS}(\xcut)
=\deltatotal\,,
\label{eq:delta_alt}
\end{equation}
for any $\nu^2\le\xcut<1$.
In particular, the value in \cref{eq:deltap_95} becomes the same as in \cref{eq:deltap_95_new} and should be used from now on for $\xcut=0.95$.

Employing \cref{eq:Bpi0eeg,eq:Bpi0ee}, the new value \eqref{eq:deltap_95_new} leads to
\begin{multline}
B\bigl(\pieegx,\,x>0.95\bigr)\\
\approx\bigl(5.82(1)_\xi+0.14\chit\bigr)\times10^{-8}\,.
\label{eq:B_SM}
\end{multline}
Other choices of $\xcut$ are possible and should be used for various cross-checks in experimental analyses.
Such results are easy to obtain employing \cref{eq:deltacut$new$} and \cref{tab:xcut}.

It is rather intriguing that the overall radiative correction found above, $\deltatotal\approx10.7\%$, is significantly larger than what one would naively expect:
Typically, one anticipates that the overall radiative corrections are $\mathcal{O}(1\%)$.
This one can see, e.g., in cases of one-photon-inclusive corrections for the $\pi^0$ Dalitz decay $\pieeg$~\cite{Joseph:1960zz,Mikaelian:1972yg,Husek:2018qdx}, $K_{e3}^+$~\cite{Cirigliano:2001mk}, or $K^+\to\pi^+e^+e^-$~\cite{Husek:inprep}.
This is closely related to the fact that at LO, the $\pi^0$ rare decay is helicity-suppressed, and this suppression is elevated by the emission of the extra photon.
The $\pieegx$ amplitude then survives the $m_e\to0$ limit, and in this sense, its overall contribution is enhanced compared to the simple $\alpha/\pi$ scaling; see also \cref{sec:BSm0} for more detailed discussion.

Finally, it seems convenient to work with $\deltam(\xcut)$ and $\deltatotal$ as the two independent objects to derive the rest of the quantities useful for experimental analysis.
For instance, the ratio of events with $x>\xcut$ to all $\pieegx$ events is then
\begin{multline}
R_+(\xcut)
\equiv\frac{B\bigl(\pieegx,\,x>\xcut\bigl)}{B\bigl(\pieegx\bigl)}\\
=\frac{1+\deltap(\xcut)}{1+\deltatotal}
=1-\frac{\deltam(\xcut)}{1+\deltatotal}\,.
\label{Bpi0eeg2all}
\end{multline}

\section{Latest measurements}
\label{sec:measurements}

The first two measurements~\cite{Fischer:1977zc,Frank:1982ea} of the $\piee$ BR led to values $\mathcal{O}\bigl(10^{-7}\bigr)$ and were, to a large extent, contradicted by the upper bound $B\bigl(\piee\bigr)\le1.3\times10^{-7}$ obtained by the third experiment~\cite{SINDRUM:1989arh}; cf.~introductions of Refs.~\cite{Deshpande:1993zn,McFarland:1993wv}.

Following the publication of the measurement at Brookhaven National Laboratory~\cite{Deshpande:1993zn}, the latest three determinations originate at Fermilab~\cite{McFarland:1993wv,KTeV:1999bjr,KTeV:2006pwx}.
Even though all the latter four are mutually compatible and listed by the latest PDG review~\cite{Workman:2022ynf}, the PDG average%
\footnote{
    One might want to approach the present PDG value carefully.
    For the listed past experiments, the PDG gather and present the quantities $B\bigl(\pieegx,\,x>0.95\bigr)$ directly as inputs for the $\piee$ BR entry.
    At the same time, PDG also provide the corrected (extrapolated) BRs quoted in the original references that used Ref.~\cite{Bergstrom:1982wk} for RCs:
    This might be misleading in light of the (significantly different) complete two-loop result~\cite{Vasko:2011pi}.
}
is largely driven by a single precise, updated measurement published by the KTeV Collaboration~\cite{KTeV:2006pwx}.
Such an unsatisfactory situation, in which only one precise measurement of the $\piee$ BR was available and in a persisting tension with the SM prediction, remained until recently.
With the long-expected update provided by NA62 in terms of a preliminary result, the circumstances are more comforting, as the new value seems nicely consistent with the SM prediction; see \cref{tab:chi}.
In the following, I summarize some interesting points regarding the two latest analyses.

\subsection{KTeV measurement}

Looking at the quantity that has been measured by KTeV~\cite{KTeV:2006pwx},
\begin{equation}
\frac{B\bigl(\pieegx,\,x>0.95\bigl)}
{B\bigl(\pieeg,\,x>0.232\bigl)}
=1.685(64)(27)\times10^{-4}\,,
\label{eq:BBKTeV}
\end{equation}
one can correctly assume that its relation with the theoretical quantity $B\bigl(\piee)$ is not simple and requires some nontrivial theoretical inputs.
The result in \cref{eq:BBKTeV} not only depends on the Dalitz-decay BR, serving here as a normalization channel, but the numerator allows for final states including soft photons with their energies limited by the lower bound on $x$.
\Cref{eq:BBKTeV} thus needs to be further processed, and it was done so in the KTeV analysis via a series of steps.%

Once the denominator is removed \{based on the NLO knowledge of the Dalitz decay, for which the radiative corrections of Ref.~\cite{Mikaelian:1972yg} were used\}, for extracting $B\bigl(\piee)$ from $B\bigl(\pieegx,\,x>0.95\bigl)$ appearing in the numerator of \cref{eq:BBKTeV}, the procedure KTeV used was the following:
\begin{enumerate}
    \item[1)] Extrapolate the radiative tail of the BR to the whole $x$ region (i.e., including the contribution of hard photons arising at low-$x$), arriving thus at the inclusive BR $B\bigl(\pieegx\bigl)$.
    \item[2)] Scale this result back by dividing by $(1+\deltatotal)$ to obtain $B\bigl(\piee\bigl)$, equivalent to the so-called ``no-radiation'' BR.
\end{enumerate}
The obtained values are listed in \cref{tab:chi}.

For the first step, the exact value of the factor $1/R_+(0.95)$ used was not quoted in Ref.~\cite{KTeV:2006pwx}, although the resulting numbers point at $R_+^\text{\cite{KTeV:2006pwx}}(0.95)\approx83.3\%$.
This is close to the value based on the exact NLO calculation and \cref{Bpi0eeg2all}:
\begin{equation}
R_+(0.95)
=\bigl[84.88(1)_\xi+0.12\chit\bigr]\%\,.
\end{equation}
For the second step, the overall radiative correction used, $\deltatotal=3.4\%$~\cite{Bergstrom:1982wk}, differs significantly from $\deltatotal$ obtained in \cref{eq:delta}.
This is due to the $\approx\!8\%$ difference in the virtual radiative corrections; cf.~\cref{eq:deltap_95} and above.

Despite the use of the up-to-date exact NLO radiative corrections, \cref{eq:BBKTeV} does not lead to a value consistent with the SM [cf.~\cref{eq:Bpi0ee}], as it might also contain additional traces of inaccurate corrections or unaccounted-for background.
One additional known issue is related to the fact that the result \eqref{eq:BBKTeV} includes the unsubtracted contribution of the interference of the $\piee$ BS with the Dalitz decay mentioned in \cref{sec:experiment}~\cite{Vasko:2011pi,Husek:2014tna,Husek:2015wta}, which is, however, numerically not that significant.%
\footnote{
    The contribution of this interference to the Dalitz decay width is negative (cf.~\cref{fig:Gamma_pi0eeg_95}), and the subtracted contribution of the Dalitz-decay events from the $x>0.95$ region is consequently overestimated.
    One could thus scale up the KTeV BR values by multiplying them by $\bigl[1-\deltap^{1\gamma\text{IR}}(0.95)\bigr]$, with $\deltap^{1\gamma\text{IR}}(0.95)\approx-0.36(1)_\chi\%$.
    Alternatively, when treating the KTeV result, $\deltap^{1\gamma\text{IR}}(0.95)$ can be added to \cref{eq:deltap_95}.
    In this regard, note that there is a misprint in Eq.~(17) of Ref.~\cite{Husek:2014tna}, as the value therein should include a negative sign.
}

\subsection{NA62 measurement}

In the NA62 experiment, the discussed radiative corrections for $\piee$ are implemented in the two NLO MC decay generators, the (trivial) 2-body and 3-body [$\pieegx$] generators.
As an important theoretical input to the analysis, the ratio of the two integrated decay widths is calculated.
It depends on $\xcut$ and $\chir$, and the uncertainty is dominated by virtual corrections.
In terms of (independent) $\deltatotal$ and $\deltam(\xcut)$, it is defined as
\begin{multline}
R_{2/3}(\xcut)
\equiv\frac{B\bigl(\pieegx,\,x>\xcut\bigr)}{B\bigl(\pieegx,\,x<\xcut\bigr)}\\
=\frac{1+\deltap(\xcut)}{\deltam(\xcut)}
=\frac{1+\deltatotal}{\deltam(\xcut)}-1\,.
\end{multline}
In the NA62 analysis,
\begin{equation}
\begin{aligned}
R_{2/3}(0.995)
&=3.034(3)_\xi+0.016\chit\,,\\
R_{2/3}(0.98)
&=4.280(4)_\xi+0.031\chit\,,
\end{aligned}
\end{equation}
with $R_{2/3}(0.995)$ being used as a primary input, and $R_{2/3}(0.98)$ is employed for systematic cross-check.
Relating $R_{2/3}(\xcut)$ with $R_+(\xcut)$ of \cref{Bpi0eeg2all}, one can write
\begin{equation}
\begin{aligned}
R_+(\xcut)
&=\frac{R_{2/3}(\xcut)}{1+R_{2/3}(\xcut)}\,,\\
R_{2/3}(\xcut)
&=\frac{R_+(\xcut)}{1-R_+(\xcut)}\,.
\end{aligned}
\end{equation}
For instance, $R_{2/3}(0.95)=5.612(5)_\xi+0.051\chit$.

The NLO Dalitz-decay generator, including the 1$\gamma$IR contribution, is also employed within the NA62 MC framework:
Its relative contribution in the $x>\xcut$ region can be obtained as
\begin{equation}
\begin{aligned}
R_\text{D}(\xcut)
\equiv\frac{B\bigl(\pieeg(\gamma),\,x>\xcut\bigr)}{B\bigl(\pieegx,\,x>\xcut\bigr)}\,,
\end{aligned}
\end{equation}
where, in the numerator, there is the Dalitz-decay width including all the NLO corrections and, in the denominator, there is the $\piee$ width including the BS.
In terms of $R(\xcut)$ of Ref.~\cite{Husek:2018qdx}, one finds
\begin{equation}
R_\text{D}(\xcut)
=\frac{R(\xcut)}{1+\deltap(\xcut)}\frac{B\bigl(\pigg\bigr)}{B\bigl(\piee\bigl)}\,.
\end{equation}
The values for $R(\xcut)$ can be read from Table III of Ref.~\cite{Husek:2018qdx}:
In particular, $R(0.95)=0.1967(35)_{a_\pi}\times10^{-8}$.
In Ref.~\cite{Husek:2018qdx}, one can also find $B\bigl(\pigg\bigr)=98.8131(6)\%$.
Employing \cref{eq:B_SM},
\begin{equation}
R_\text{D}(0.95)
=\bigl[3.34(6)_{a_\pi}-0.08\chit\bigr]\%\,.
\end{equation}

\section{Relation to the Dalitz decay}
\label{sec:1gIR}

\begin{figure}
\centering
\begin{subfigure}{0.32\columnwidth}
    \centering
    \includegraphics[width=\linewidth]{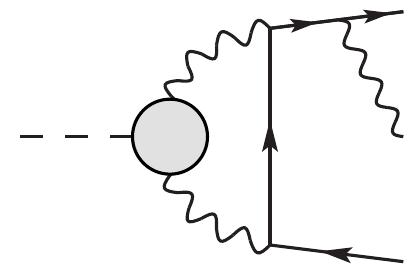}
    \caption{}
\label{fig:BS_a}
\end{subfigure}
\hfill
\begin{subfigure}{0.32\columnwidth}
    \centering
    \includegraphics[width=\linewidth]{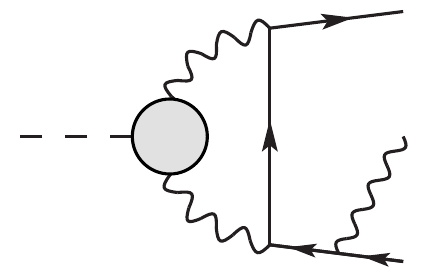}
    \caption{}
\label{fig:BS_b}
\end{subfigure}
\hfill
\begin{subfigure}{0.32\columnwidth}
    \centering
    \includegraphics[width=\linewidth]{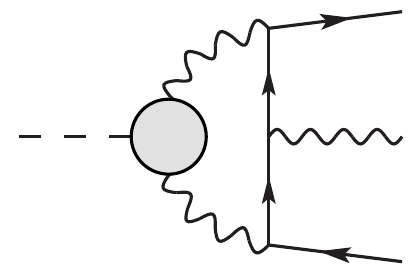}
    \caption{}
\label{fig:BS_c}
\end{subfigure}
\caption{
    The NLO bremsstrahlung contribution to $\piee$, also known as the two-photon-exchange or one-photon-irreducible (1$\gamma$IR) contribution to $\pieeg$ at one loop: photon emission from the external [(a), (b)] and internal (c) charged-lepton lines.
}
\label{fig:BS}
\end{figure}

Since the BS correction to $\piee$ embodies the same final state as the Dalitz decay of the neutral pion, their interference should be considered in calculating radiative corrections to the latter.%
\footnote{
Let me remind the reader that this interference is subtracted together with the Dalitz decay itself in the NA62 $\piee$ BR measurement so that the studied (inclusive) quantities are only derived from $\piee$, with no Dalitz-decay contribution.
}
This contribution was historically considered negligible, assuming a straightforward relation of the $\piee$ BS diagrams (see \cref{fig:BS}) with the LO:
It was believed that these are also helicity-suppressed~\cite{Mikaelian:1972yg}.
Further iterations justified these assumptions~\cite{Lambin:1985sb} based on Low's theorem~\cite{Low:1958sn}.
Only explicit calculation in the $m_e\to0$ limit~\cite{Tupper:1983uw,Tupper:1986yk} showed a nonvanishing result and faults in the preceding reasoning.
In Ref.~\cite{Tupper:1986yk}, it was also pointed out that one cannot simply interchange the $m_e\to0$ and $k\to0$ limits, with $k$ being the photon four-momentum.
A detailed discussion and additional arguments based on the nonanalyticity of the amplitude can be found in Secs.~3.2 and 3.3 of Ref.~\cite{Kampf:2005tz}.

In this section, I first examine the result in the former, massless limit, obtained in the approximation in which the $\pi^0$ is pointlike.
In this case, the correction to the doubly differential Dalitz decay width in variables $x$ and 
\begin{equation}
y
=\frac{2P_{\pi^0}\cdot(p_{e^+}-p_{e^-})}{\Mpi^2(1-x)}
\label{eq:y}
\end{equation}
is simply~\cite{Tupper:1986yk}
\begin{equation}
\deltax^{1\gamma\text{IR}}(x,y)
=-\frac\alpha\pi
\biggl\{
\frac{x}{(1-x)(1+y^2)}+G(x_+,x_-)
\biggr\}\,,
\label{eq:delta_1gIR}
\end{equation}
with $x_\pm=\frac2{\Mpi^2}k\cdot p_{e^{\pm}}=\frac12(1-x)(1\pm y)$ and
\begin{multline}
G(x_1,x_2)\\
\equiv\log(x_1)\log(x_2)+\operatorname{Li}_2(1-x_1)+\operatorname{Li}_2(1-x_2)-\frac{\pi^2}6\,.
\label{eq:G}
\end{multline}
\{Note that I use here slightly different notation than in Eq.~(15) of Ref.~\cite{Tupper:1986yk} (there is also an obvious typo).\}.
The correction to the $x$ spectrum reads~\cite{Tupper:1986yk}
\begin{multline}
\deltax^{1\gamma\text{IR}}(x)
=\frac\alpha\pi
\biggl\{
-\log^2(1-x)+
\frac{2x}{1-x}\biggl[
\frac{1-\log(1-x)}{1-x}\\
-\frac58
+\bigg(1+\frac{1+x}{2(1-x)^2}\bigg)
\bigg(\operatorname{Li}_2x-\frac{\pi^2}6\bigg)
\biggr]
\biggr\}\,.
\label{eq:delta_1gIRx}
\end{multline}
One can easily obtain the behavior at $x\to0$, arriving at
\begin{multline}
\deltax^{1\gamma\text{IR}}(x)\big|_{x\to0}
=\frac\alpha\pi\biggl[
-\frac14\bigl({2\pi^2-3}\bigr)x\\
-\frac14\bigl({4\pi^2-27}\bigr)x^2
+\mathcal{O}\big(x^3\big)
\biggr]\,,
\label{eq:delta1gIRx}
\end{multline}
i.e., both the first and second derivatives are negative, and at $x\to1$,
\begin{multline}
\deltax^{1\gamma\text{IR}}(x)\big|_{x\to1}
=\frac\alpha\pi\biggl[
-\frac34\frac1{1-x}
-\log^2(1-x)\\
+\frac{13}6\log(1-x)
+\mathcal{O}\big((1-x)^0\big)
\biggr]\,,
\end{multline}
which shows that the 1$\gamma$IR contribution to decay width approaches zero by a factor of $(1-x)$ slower than the Dalitz-decay width proportional to $(1-x)^3$, leading to large corrections to the tail; see also \cref{fig:Gamma_pi0eeg_95}.

As the effect of radiative corrections on the TFF slope estimate can be written as $(1+a_{+\text{QED}}x)^2=(1+ax)^2[1+\deltax(x)]$, with $a$ being the purely hadronic quantity, this means that, taking first derivative as $x\to0$,
\begin{equation}
\Delta a
\equiv a_{+\text{QED}}-a
=\frac12\frac{\mathrm{d}\deltax(x)}{\mathrm{d}x}\bigg|_{x\to0}+a\lim_{x\to0}\deltax(x)\,.
\end{equation}
As the second term on the right-hand side is of the size of NNLO corrections and can be neglected, one can read off from \cref{eq:delta1gIRx} the correction brought in by the $1\gamma$IR contribution (note also that $\deltax^{1\gamma\text{IR}}(0)=0$ anyway), $\Delta a|_{1\gamma\text{IR}}=-\frac\alpha\pi\frac18\bigl({2\pi^2-3}\bigr)\approx-0.486\%$; 
the exact value employing physical $m_e$ would correspond to $\Delta a|_{1\gamma\text{IR}}=-0.502(7)\%$.
As the intuitive size of the slope is $a\approx3\%$, such a correction is not negligible.
Note that an estimate for $\Delta a$ based on the total NLO correction is $\Delta a\approx-6.3\%$, about twice the size of $a$, which makes the knowledge of RCs essential:
An expected value for the slope $a_{+\text{QED}}$ extracted without considering RCs would thus be of reasonable size, but opposite sign.

The approximate results (\ref{eq:delta_1gIR}--\ref{eq:delta_1gIRx}) can be checked against the interference of the complete result for the $\pieegx$ amplitude of Ref.~\cite{Husek:2014tna} and the LO Dalitz-decay amplitude.
And vice versa:
They serve as a basic check of the complete result.
These incredibly simple expressions for $m_e\to0$ turn out to approximate the exact result very well.
Having the complete $\pieegx$ amplitude at hand for comparison, taking the pion here as pointlike is also justified as the terms proportional to $\chir(\mu)$ are $\mathcal{O}\big(m_e^2\big)$.
Recall that this is not the case for the $\piee$ amplitude, which would be UV-divergent for pointlike $\pi^0$ and for which the model-dependence [on the parameter $\chir(\mu)$] is rather substantial.

\section{The rare-decay bremsstrahlung in the massless limit}
\label{sec:BSm0}

With the results of the previous section at hand, one can obtain rather easily an analytical expression for the $\pieegx$ decay width in the limit $m_e\to0$; for the numerical result, see \cref{fig:Gamma_pi0eeg}.
In general, the $\pieegx$ amplitude reads (four-momenta of electron, positron, and photon are denoted $p$, $q$, and $k$, respectively)
\begin{multline}
i\mathcal{M}^\text{BS}(p,q,k)\\
=-\frac{i e^5}{2}\mathcal{F}_{\pi^0\gamma\gamma}^\text{LO}\epsilon^{*\rho}(k)\,
\bar{u}(p,m)\Gamma_\rho(p,q,k)\gamma_5 v(q,m)\,,
\label{eq:M1gI}
\end{multline}
with $\Gamma_\rho$ that can be written in terms of three subamplitudes $P$, $A$, and $T$ in a manifestly gauge-invariant form as~\cite{Husek:2014tna,Husek:2015sma,Kampf:2005tz}
\begin{equation}
\begin{aligned}
\Gamma_\rho(p,q,k)
&=P(x,y)\bigl[(k\cdot p) q_\rho-(k\cdot q) p_\rho\bigr]\\
&+A(x,+y)\bigl[\gamma_\rho(k\cdot p)-p_\rho\slashed k\bigr]\\
&-A(x,-y)\bigl[\gamma_\rho(k\cdot q)-q_\rho\slashed k\bigr]\\
&+T(x,y)\,\gamma_\rho\slashed k\,.
\end{aligned}
\end{equation}
Up to an overall phase factor, the subamplitude $A$ relates to $A_i$ of Ref.~\cite{Tupper:1986yk} as $A_1=A(x,y)$, $A_2=A(x,-y)$, and $A_3=-k\cdot p\,A_1+k\cdot q\,A_2$.
As the contributions involving subamplitudes $P$ and $T$ vanish in the $m_e\to0$ limit, the matrix element squared in this limit reduces to
\begin{multline}
\overline{\big|\mathcal{M}_{m_e\to0}^\text{BS}(x,y)\big|^2}
=\alpha^3\Mpi^3\Gamma_{\pi^0\to\gamma\gamma}^\text{LO}\\
\times x(1-x)^2
\Bigl[
 (1-y)^2\bigl|16\pi^2A(x,y)\bigr|^2
 +(y\to-y)
\Bigr]\,,
\label{eq:M2BSm0}
\end{multline}
with
\begin{equation}
-16i\pi^2A(x,y)
=-\frac4{\Mpi^2}\biggl[\frac1{(1-x)(1-y)}+\frac1x\,G(x_+,x_-)\biggr]
\label{eq:Axy}
\end{equation}
and $G$ the same as in \cref{eq:G}.
This expression has been checked explicitly using the exact result and is numerically very close to setting $P$ and $T$ to zero in the complete expression with physical $m_e$.
As another cross-check, once the $A(x,y)$ from \cref{eq:Axy} is plugged into Eq.~(26) of Ref.~\cite{Husek:2015sma}, one recovers \cref{eq:delta_1gIR}.
The function $G$ corresponds, up to overall factors, to the regular finite part of the massless IR-divergent box diagram \{$G$ is equivalent to the second line of Eq.~(I.12) of Ref.~\cite{Bern:1993kr}\}.

In general, for a gauge-invariant $\pieegx$ amplitude, one needs contributions from all three diagrams in \cref{fig:BS}.
However, it is intriguing to explore their very different properties.
The diagrams in \cref{fig:BS_a,fig:BS_b} are UV-divergent in the pointlike-$\pi^0\gamma\gamma$ approximation (and thus require counterterms in the sense of the expansion in \cref{fig:LO}), lead to an IR-divergent decay width, and vanish in the massless limit.
In the soft-photon limit, their sum alone is already gauge-invariant, as the ``box'' diagram of \cref{fig:BS_c} vanishes in that limit.
Furthermore, on the other hand, the contribution of the box is UV- and IR-finite.%
\footnote{
One subtlety may be emphasized at this point.
The RCs presented in this work do not consider any dependence on a TFF model related to the diagram in \cref{fig:BS_c}, and only its pointlike form is used.
As mentioned at the end of Sec.~4 of Ref.~\cite{Husek:2017vmo}, which suggests a framework for calculating the diagrams in \cref{fig:BS} for rational TFFs, such corrections are expected to be of order $m_e^2/M_\rho^2$ and $\Mpi^2/M_\rho^2$ and can be treated as negligible in the global context.
}

It is an important observation that, in general, only the box diagram in \cref{fig:BS_c} contributes to the amplitude $A$ (and it also contributes to $P$ and $T$).
As the subamplitude $A$ is the only one contributing in the massless limit, the diagrams with radiation from the external legs do not contribute at all in this limit.
In particular, for the (small) physical electron mass, the dominant contribution to $\deltax^{1\gamma\text{IR}}(x,y)$ thus comes from the box diagram: Here, $P$ does not contribute, and $T$'s contribution is suppressed; see Eq.~(26) of Ref.~\cite{Husek:2015sma}.
In contrast, the $P$ subamplitude dominates in the soft-photon limit and carries the potential IR-divergent behavior.

To summarize, one can thus see the different properties of the $\pieegx$ amplitude in the $m_e\to0$ and $k\to0$ limits manifestly at subamplitude level.
Respectively, only one subamplitude ($A$ or $P$) survives each limit ($m_e\to0$ or $k\to0$), a different one in each case.
The original argument is thus valid: The BS contribution to $\piee$ can be approximated, for $x\to1$, by its soft-photon form, in which the amplitude factorizes and is proportional to the LO amplitude, which in turn is proportional to $m_e$ and thus vanishes in the massless limit.
Such a contribution to the $\pi^0$-Dalitz RCs would be negligible, although, as described above, it is related to $P$, which does not actually contribute to $\deltax^{1\gamma\text{IR}}$.
However, there are subleading terms in the soft-photon limit, gauge-invariant on their own and thus not restricted by Low's theorem, that become dominant away from $x\to1$ and survive the massless limit.
They originate in the box diagram that is (in general) the only one contributing to $A$, which in turn dominates the nonnegligible contribution to $\deltax^{1\gamma\text{IR}}$.

\section{Summary}

In light of the newest BR measurement of the $\piee$ decay being finalized by the NA62 Collaboration, reviewing the status of the associated RCs seemed appropriate.
It turned out that some minor numerical updates were to take place.
The range of presented results broadened here compared to Ref.~\cite{Husek:2014tna} which first evaluated the exact $\pieegx$ amplitude, meaning the NLO BS process for $\piee$ depicted in \cref{fig:BS}, calculated beyond the soft-photon approximation that was used earlier in Ref.~\cite{Vasko:2011pi}.

The theoretical $\piee$ BR can be written in a model-independent way as
\begin{equation}
B\bigl(\piee\bigr)
\approx\bigl(6.19+0.15\chit\bigr)\times10^{-8}\,.
\label{eq:Bpi0ee2}
\end{equation}
In this work, a generous range for SM-compatible values is considered, assuming a conservative interval $\chit\in(-1,1)$;
see the discussion around \cref{eq:Bpi0ee}.
The studied radiative corrections relate $B\bigl(\piee\bigr)$ with experimental results.
The most important quantity in this regard, $\deltap(\xcut)$, has been updated, and the value [cf.~\cref{eq:deltap_95_new}]
\begin{equation}
\deltap(0.95)
=\bigl[-6.06(7)_\xi-0.08\chit\bigr]\%
=-6.1(2)\%
\end{equation}
should be used from now on for experimental results that consider $\xcut=0.95$.
In particular, one thus expects [reproducing \cref{eq:B_SM}]
\begin{multline}
B\bigl(\pieegx,\,x>0.95\bigr)\\
\approx\bigl(5.82(1)_\xi+0.14\chit\bigr)\times10^{-8}\,.
\end{multline}
For any other cut-off, one can employ \cref{eq:deltacut$new$}, combined with the knowledge of the overall correction $\deltatotal$ from \cref{eq:delta} and the values for $\deltam(\xcut)$ from \cref{tab:xcut}; cf.~\cref{fig:delta_pm}.

The KTeV measurement triggered an enlarged interest in this decay as their result differed considerably from the SM prediction.
As the new NA62 result is, within available precision, compatible with our understanding based on the SM considerations, as the next step, one can aim to extract information on the transition form factor.
The present experimental error is twice as large as the uncertainty band indicated in \cref{eq:Bpi0ee2} and extraction of $\chir$ is at the moment inconclusive; cf.~\cref{tab:chi}.
In this regard, the current (NLO) knowledge of radiative corrections is expected to become a limitation only for experimental analyses delivering results with precision well below the percent level.

Theoretical inputs represent indispensable ingredients to experimental analyses.
Not only do the radiative corrections need to be considered consistently in each MC sample, but their relative weights need to be known so that they can be combined and the resulting simulated spectra compared with data.
For the rare and Dalitz $\pi^0$ decays, the necessary quantities can be calculated precisely in terms of numerical integrals; for more complicated processes, one might need to invoke MC methods.
Some important inputs, the ratios introduced here as $R_+$, $R_{2/3}$, or $R_\text{D}$, are presented in \cref{sec:NLO,sec:measurements}.

The quantities given in this work explicitly state their dependence on the low-energy parameter $\chit$.
This is potentially useful to obtain more precise values for these quantities when a particular value of $\chir$ is considered or known.
With no precise experimental knowledge of $\chir$, the presented results allow for interpreting the $\chit$ part as a systematic uncertainty.
For instance, varying $\chit\in(-1,1)$ thus leads to rather conservative estimates that can be considered model-independent.

In the last part of the paper, the relation with the Dalitz decay was discussed.
For several decades, the importance of the two-photon exchange (1$\gamma$IR) contribution to the Dalitz-decay radiative corrections has been repeatedly questioned due to Low's theorem and the helicity-suppressed LO $\piee$ amplitude.
As it was reminded in \cref{sec:1gIR}, the $\pieegx$ amplitude does not vanish in the massless limit, and a very simple analytical formula [\cref{eq:delta_1gIR}] can be written down in this case, driven by a single diagram shown in \cref{fig:BS_c}.
Misleading outcomes thus arise when the order in which the massless and soft-photon limits are taken is not distinguished --- like, in particular, in the simplest case when assuming that the complete amplitude is helicity-suppressed since it is the case in the soft-photon limit.

Following up on the elegant result \eqref{eq:delta_1gIR} and identifying its relation to the $A$ subamplitude in \cref{sec:BSm0}, the exact $\pieegx$ amplitude can be very well approximated by taking $A$ in the massless limit and $P$ in the soft-photon limit, in which, on the other hand, the diagrams in \cref{fig:BS_a,fig:BS_b} are dominant.
Summing simply the widths stemming from the results in \cref{eq:M2BSm0,eq:deltax}, such approximation works for the entire range of $x$, although the difference is most prominent in the soft-photon region contributing to the experimental BR; cf.~\cref{fig:Gamma_pi0eeg}.

\begin{acknowledgments}

I thank Michal Kovaľ for his valuable comments on the manuscript and Karol Kampf and Jiří Novotný for encouraging discussions.

This work was supported by the MSCA Fellowships CZ--UK2 project No.~CZ.02.01.01/00/22\_010/0008115 financed by Programme Johannes Amos Comenius (OP JAK).

\end{acknowledgments}


\appendix

\section{Soft-photon bremsstrahlung}
\label{sec:BS}

The $\pieegx$ amplitude squared can be written in soft-photon approximation as
\begin{equation}
\overline{\big|\mathcal{M}^\text{soft}(k,p,q)\big|^2}
=e^2\Deltasoft(k,p,q)\overline{\big|\mathcal{M}^\text{LO}(p,q)\big|^2}\,,
\label{eq:Msoft}
\end{equation}
where $\mathcal{M}^\text{LO}$ is the leading-order amplitude in the QED expansion represented in \cref{fig:LO} and leading (up to $\DeltaZ$) to \cref{eq:pieeLO}.
As in \cref{sec:BSm0}, four-momenta of electron, positron, and photon are denoted $p$, $q$, and $k$, respectively.
In kinematical variables $x$ and $y$ of \cref{eq:x,eq:y}, $\Deltasoft(k,p,q)$ becomes
\begin{equation}
\Deltasoft(k,p,q)
\equiv
-\bigg(\frac{p^\rho}{k\cdot p}-\frac{q^\rho}{k\cdot q}\bigg)^2
=\frac{16x}{\Mpi^2}\frac{1-\frac{\nu^2}x-y^2}{(1-x)^2(1-y^2)^2}\,.
\end{equation}
Including phase-space factors $\sqrt{1-\nu^2}/(16\pi\Mpi)$ and $\Mpi(1-x)/(8\pi)^3$ for $\piee$ and $\pieeg$, respectively, one gets for the decay widths ratio
\begin{equation}
\begin{aligned}
\deltax^\text{(soft)}(x,y)
&=\frac{\mathrm{d}\Gamma_{\pieegx}^\text{(soft)}(x,y)}{\mathrm{d}x\,\mathrm{d}y}\bigg/\Gamma_{\piee}^\text{LO}\\
&=\frac\alpha\pi\frac{\Mpi^2(1-x)}{8\sqrt{1-\nu^2}}\,\Deltasoft(x,y)\,.
\end{aligned}
\label{eq:delta_soft_xy}
\end{equation}
However, combining the expressions from Ref.~\cite{Husek:2014tna} [Eqs.~(7--8), (23--24), and (A.14--A.16)], one finds that
\begin{equation}
\deltax^\text{(soft)}(x,y)
=\frac1x\deltax_\text{\cite{Husek:2014tna}}^\text{(soft)}(x,y)\,.
\label{eq:delta_soft_xy_2014}
\end{equation}
This is connected to the fact that, in Ref.~\cite{Husek:2014tna}, a general prescription for the matrix element squared [Eq.~(23)] for the process $\pieeg$ was used that includes a square of the pseudoscalar Dirac structure leading to $2(p+q)^2=2\Mpi^2x$, whereas in the standard soft-photon-BS approach [\cref{eq:Msoft}], one factorizes the LO matrix element squared in which $(p+q)^2=\Mpi^2$.
As it was done in Ref.~\cite{Husek:2014tna}, plugging the subamplitude $P$ in the soft-photon limit $x\to1$ ($P_\text{soft}$) into the complete expression does not give the equivalent answer with respect to the treatment employed in Ref.~\cite{Vasko:2011pi}, and the compensatory term does not have the required meaning.
As shown below, removing a factor of $x$ is not enough to arrive at a consistent answer.

As the $\piee$ process does not depend on $y$ (nor $x$), one can directly integrate $\deltax^\text{(soft)}(x,y)$ of \cref{eq:delta_soft_xy} over the angular variable $y\in\bigl(-\beta(x),\beta(x)\bigr)$, with $\beta(x)\equiv\sqrt{1-\nu^2/x}$, which leads to
\begin{equation}
\deltax^\text{(soft)}(x)
=\frac{c_1(x)}{1-x}\,,
\label{eq:deltax}
\end{equation}
with
\begin{equation}
c_1(x)
\equiv\frac\alpha\pi\frac{(-2)x\beta(x)}{\sqrt{1-\nu^2}}\biggl[1+
\frac{1+\beta^2(x)}{2\beta(x)}
\log\frac{1-\beta(x)}{1+\beta(x)}\biggr]\,.
\end{equation}
To integrate this further over $x\in(\xcut,1)$, one must be cautious and regularize the integrand first.
Moreover, in the soft-photon-limit approach, one only integrates over photon energies and emission angles, keeping the rest of the particles intact.
This effectively means taking $x\to1$ whenever possible, i.e., replacing $\beta(x)\to\beta=\sqrt{1-\nu^2}$ in \cref{eq:deltax}, which leads to
\begin{equation}
\deltax_{x\to1}^\text{(soft)}(x)\\
=\frac{c_1(1)}{1-x}\,,
\label{eq:deltax1}
\end{equation}
with the same $c_1(1)=c_1$ as in \cref{eq:ci}.
The regularized result for the (equivalent) integral (keeping $p$ and $q$ fixed by the 2-body kinematics)
\begin{equation}
\deltap^\text{(soft)}(E_\text{max})
=e^2
\hspace{-2mm}
\int\limits_{E_k<E_\text{max}}
\hspace{-2mm}
\frac{\mathrm{d}^3k}{(2\pi)^32E_k}\,\Deltasoft(k,p,q)
\end{equation}
can be written as
\begin{multline}
\deltap^\text{(soft)}(E_\text{max})
=\frac\alpha\pi
\biggl\{
\biggl(-2\log\frac{2E_\text{max}}{\Mpi}+\widetilde X\biggr)
\bigl(1+\widetilde\beta\log z\bigr)\\
-\frac1\beta\log z-\widetilde\beta\,
\biggl[
\frac12\log^2z
+2\text{Li}_2(1-z)
\biggr]
\biggr\}\,.
\label{eq:deltapBS}
\end{multline}
For the photon-mass ($\Lambda$) regularization scheme, one then has $\widetilde X_\Lambda=-\log\frac{\Mpi^2}{m_e^2}-\log\frac{m_e^2}{\Lambda^2}$, and for the dimensional-regularization scheme used in Ref.~\cite{Vasko:2011pi}, one finds $\widetilde X_\epsilon=-\log\frac{\Mpi^2}{m_e^2}+\frac1\epsilon-\gamma_\mathrm{E}+\log4\pi$.
Considering that in the assumed frame (pion rest frame), $E_\text{max}={\Mpi}\frac12(1-\xcut)$, and that the IR divergences (terms proportional to $-\log\frac{m_e^2}{\Lambda^2}$ or $\frac1\epsilon$) cancel against the virtual-correction contributions, one can easily read off $c_1$ and $c_2$; cf.~\cref{eq:ci}.

Now, the (IR-finite) term $\Deltaxcut^\text{BS}(\xcut)$ that compensates for the use of the soft-photon limit becomes well-defined once the divergences are subtracted already at integrand level,
\begin{equation}
\begin{aligned}
\Deltaxcut^\text{BS}(\xcut)
&=\int_{\xcut}^1\deltax^\text{(BS)}(x)-\deltax^\text{(soft)}(x)\,\mathrm{d}x\,,
\end{aligned}
\label{eq:DeltaBsxcut}
\end{equation}
with $\deltax^\text{(BS)}(x)$ stemming from the square of the exact $\pieegx$ amplitude of Ref.~\cite{Husek:2014tna}.
To properly match this with Ref.~\cite{Vasko:2011pi} [to obtain input for \cref{eq:delta_alt}], one should use $\deltax^\text{(soft)}(x)=\deltax_{x\to1}^\text{(soft)}(x)$ of \cref{eq:deltax1}
in the above integrand.
In other words, then the integral \eqref{eq:DeltaBsxcut} correctly pairs with the BS in the soft-photon approximation $\deltap^\text{(soft)}(\xcut)$ of \cref{eq:deltapBS} entering the correction $\deltap^\text{\cite{Vasko:2011pi}}(\xcut)$ in \cref{eq:delta$x_cut$soft}.
Naturally, the exact case $\deltax^\text{(BS)}(x)$ is best approximated by $\deltax^\text{(soft)}(x)$ from \cref{eq:deltax}.
It also has the correct behavior as $x\to\nu^2$, for which it vanishes.
Finally, in Ref.~\cite{Husek:2014tna}, as mentioned above, $x\deltax^\text{(soft)}(x,y)$ was instead used within $\Deltaxcut^\text{BS}(\xcut)$.
See also \cref{fig:DeltaBS} for a comparison of these approaches.

Employing the expressions above, it is straightforward to find the endpoints of the lines in \cref{fig:DeltaBSx} [using $\deltax^\text{(soft)}(x)$ as it corresponds to $\deltax^\text{(BS)}(x)$ at these points]:
\begin{equation}
\begin{aligned}
\bigl[\deltax^\text{(soft)}(x)-\deltax_{x\to1}^\text{(soft)}(x)\bigr]_{x\to\nu^2}
&=-c_1/(1-\nu^2)\approx-4.72\%\,,\\
\bigl[\deltax^\text{(soft)}(x)-\deltax_{x\to1}^\text{(soft)}(x)\bigr]_{x\to1}
&=-c_1'(x)\big|_{x\to1}\approx-5.18\%\,,\\
\bigl[\deltax^\text{(soft)}(x)-\deltax_\text{\cite{Husek:2014tna}}^\text{(soft)}(x)\bigr]_{x\to1}
&=c_1\approx4.72\%\,.
\end{aligned}
\end{equation}


\renewcommand{\raggedright}{}

\providecommand{\href}[2]{#2}\begingroup\raggedright\endgroup

\end{document}